\documentclass[12pt]{JHEP3}

\newcommand{\bea}{\begin{eqnarray}}
\newcommand{\beal}[1]{\begin{eqnarray}\label{#1}}
\newcommand{\eea}{\end{eqnarray}} 
\newcommand{\be}{\begin{equation}} 
\newcommand{\bel}[1]{\begin{equation}\label{#1}}
\newcommand{\ee}{\end{equation}}

\newcommand{\bit}{\begin{itemize}}
\newcommand{\eit}{\end{itemize}}
\newcommand{\ben}{\begin{enumerate}}
\newcommand{\een}{\end{enumerate}}

\newcommand{\req}[1]{(\ref{#1})}

\def\d{\partial}
\def\half{\frac{1}{2}}

\def\alp{\leavevmode\ifmmode {\alpha^\prime} \else ${\alpha^\prime}$ \fi}
\def\mps{M_P^2}

\title{On Power Law Inflation in DBI Models}

\preprint{}

\author{Micha\l\ Spali\'nski\footnote{Email: mspal@fuw.edu.pl}\\
So\l tan Institute for Nuclear Studies\\
ul. Ho\.za 69, 
00-681 Warszawa, Polska.
}

\abstract{
Inflationary models in string theory which identify the inflaton with an
open string modulus lead to effective field theories with non-canonical
kinetic terms -- Dirac-Born-Infeld scalar field theories. 
In the case of a $D$-brane moving in an AdS throat with a quadratic scalar
field potential DBI kinetic terms allow a novel realization of
power law inflation. This note adresses the
question of whether this 
behaviour is special to this particular choice 
of throat geometry and potential. The answer is that for any
throat geometry one can explicitly find a potential which leads to power
law inflation. This generalizes the well known fact that an exponential
potential gives power law inflation in the case of canonical kinetic
terms. 
}


\begin{document}

\section{Introduction}

The possibility that the inflaton might be on open string
modulus \cite{Dvali:1998pa,Kachru:2003sx} has
attracted much attention over the past couple of 
years (for reviews and references see
\cite{Quevedo:2002xw,Cline:2006hu,Kallosh:2007ig,HenryTye:2006uv}).    
This scenario identifies the inflaton
with the position of a mobile $D$-brane 
moving on a compact $6$-dimensional submanifold of spacetime. 
This option is
interesting in that the effective field theory is rather distinct from most
inflationary scalar field theories considered before. It is well
motivated by string computations \cite{Leigh:1989jq} and involves non-polynomial kinetic
terms of the Dirac-Born-Infeld type which impose a maximum speed on the
itinerant $D$-brane \cite{Silverstein:2003hf}. The effective action also
involves another function of 
the scalar field besides the potential. This function (denoted below by
$f$) encodes the local geometry of the compact manifold traversed by the 
$D$-brane.  

One consequence of this scenario is the
possibility of a novel realization\footnote{Other realizations of
  power law inflation in string theory are described in
  \cite{Becker:2005sg} and \cite{Dimopoulos:2005ac}.} 
of power law of inflation
\cite{Lucchin:1984yf}. In scalar  
field 
theories with canonical kinetic terms power law inflation is realized by an
exponential potential\footnote{
In brane inflation models power law inflation has a natural conclusion in
brane annihilation, following which the energy released is 
transferred to standard-model degrees of freedom
\cite{Barnaby:2004gg}--\cite{Chialva:2005zy} (see 
also \cite{Allahverdi:2006iq}, \cite{Allahverdi:2007wh}). 
}. In the scenario considered in \cite{Silverstein:2003hf} such
behaviour was shown to arise from a quadratic potential 
in the specific case when the function $f$ describes an anti-de Sitter
throat in the local geometry \cite{Giddings:2001yu}. This solution is valid in
the ``ultra-relativistic'' approximation, which is characterized by the 
inflaton expectation value saturating the moduli space speed limit. 

It is natural ask to what extent this result 
is tied to the particular choice of throat geometry and potential
made in \cite{Silverstein:2003hf}. 
The answer, as shown below, is that this type of solution exists
whenever the throat 
geometry and potential are related in a specific way. 
The basic observation is that since the
scale factor behaves as a power of cosmic time, 
the equation state must be of the baryotropic type $p=w\rho$ with constant 
$w$. 
Indeed, for FLRW models with matter in the form of a perfect fluid the
Einstein equations reduce to 
\bea
\dot{\rho} &=& - 3 H (p+\rho) \label{conserv}\\
3\mps H^2&=& \rho \label{friedman} \ ,
\eea
where $M_P$ is the reduced Planck mass ($M_P^2=1/8\pi G$), the dot indicates
a time derivative and $H\equiv \dot{a}/a$.   
If the equation of state takes the baryotropic form $p=w\rho$ with constant
$w$ then equations \req{conserv} and \req{friedman} can be integrated
and yield 
\bel{scalfact}
a(t) = a_0 (t-t_0)^{\frac{2}{3(w+1)}} \ ,
\ee
where $a_0$ is a constant. In particular, power law inflation appears for
$w<-1/3$.  Conversely, if a solution for the scale factor of the above form
is assumed, then one can easily show that the ratio $p/\rho$ is
constant. This is true independently of how the pressure and energy density
are expressed in terms of the scalar field. 

Given this, one can ask what form the effective field theory needs to take
in order that the ratio $p/\rho$ be constant. 
In the case of canonical kinetic terms the answer is
simple to determine and well known -- the potential has to be an
exponential of the scalar field. For DBI kinetic terms it appears that one
cannot give an analytic solution in general, however in the
``ultra-relativistic'' regime the problem simplifies greatly and one finds a
one-parameter family of potentials (for a given $f$), 
of which the case found in \cite{Silverstein:2003hf} is a member. 

This note is organized as follows: 
Section \ref{canonsect} 
discusses, in case of canonical kinetic terms, how 
the potential can be determined for a given equation of state by solving a
simple differential equation. This approach is then applied to DBI scalar field
theories in section \ref{dbisect}. In this case the 
differential equation cannot be solved in general, but in the
``ultra-relativistic'' limit an analytic solution can easily be given. The
relation of this solution to the one found in \cite{Silverstein:2003hf} is discussed.

\section{Canonical Kinetic Terms}
\label{canonsect}

In the canonical case the effective action for the inflaton is of the
form 
\be
S = - \int d^4 x \sqrt{-g} \Big(\half (\d\phi)^2 + V(\phi)\Big) \ .
\ee
For spatially homogeneous field configurations this leads to field
equations \req{conserv}, \req{friedman} with 
\bea
p     &=&\half \dot{\phi}^2 - V(\phi) \label{canonp}\\
\rho  &=&\half \dot{\phi}^2 + V(\phi) \label{canonrho} \ .
\eea
For a given potential one can always write an equation of the form
$p=w\rho$, but 
in general $w$ depends on the scalar field. The converse is also possible:
given a function $w(\phi)$ one can find the corresponding potential
$V(\phi)$. One way to do this uses the Hamilton-Jacobi formalism,
which is also very convenient in the DBI case. The field equations can be
written in first order
form \cite{Markov:1988yx,Muslimov:1990be,Salopek:1990jq,Kinney:1997ne}, 
treating $\phi$ as the evolution parameter in place of $t$. 
From \req{conserv}, \req{friedman} and \req{canonp}, \req{canonrho} it
follows that   
\bel{dotphi}
\dot{\phi} = -2 M_P^2 H'(\phi) \ ,
\ee
where the prime denotes a derivative with respect to $\phi$. Using this in
\req{friedman} gives 
\bel{hjcan}
3\mps H^2 - V = 2 M_P^4 H'^2 \ .
\ee
This is the Hamilton-Jacobi form of the field
equations \cite{Markov:1988yx,Muslimov:1990be,Salopek:1990jq,Kinney:1997ne}.   

The idea now is to find a closed equation for $H(\phi)$ which captures the 
condition $p=w\rho$. At the outset this can be regarded as a
parameterization of a family of equations of state with some prescribed
dependence of $w$ on on $\phi$, specializing to constant $w$ at the
end. From \req{canonp} - \req{dotphi} one obtains
\be
p = 4 M_P^4 H'^2 - \rho \ ,
\ee
so the equation $p=w\rho$ becomes (using \req{friedman})
\bel{barcan}
4\mps H'^2 = 3 (w+1) H^2 \ .
\ee
This simple differential equation for $H(\phi)$ is easily solved and the
result is 
\be
H(\phi) = H_0 \exp \Big(
\pm \frac{\sqrt{3}}{2 M_P} \int d\phi \sqrt{w(\phi) + 1}
\Big) \ ,
\ee
where $H_0$ is an integration constant. 
The corresponding potential is easily obtained from \req{hjcan}:
\be
V(\phi) = \half (1-w(\phi)) H(\phi)^2 \ .
\ee
In the special case of constant $w$ one obtains an exponential
potential: 
\be
V(\phi) = \half(1-w) H_0^2 \exp \Big(- \sqrt{3(w + 1)}\
\frac{\phi}{M_P} \Big) \ .
\ee
Thus one recovers the well known result of Lucchin and
Matarrese \cite{Lucchin:1984yf}. This simple calculation is straightforward
to repeat in the case of DBI kinetic terms.

\section{DBI Kinetic Terms}
\label{dbisect}

The effective action for the inflaton is
the Dirac-Born-Infeld action, which for spatially homogeneous inflaton
configurations takes the form \cite{Silverstein:2003hf} 
\bel{dbi}
S = - \int d^4x\ a(t)^3\ \Big(f(\phi)^{-1}(\sqrt{1-f(\phi) \dot{\phi}^2}-1) +  
V(\phi)\Big)  \ .
\ee
The function $f$ appearing here can be expressed in terms of the warp
factor in the metric and the $D$3-brane tension. The function $f$
appearing here is positive by construction\footnote{Similar actions with
  negative  $f$ have also been discussed in the
  literature\cite{Mukhanov:2005bu}--\cite{Babichev:2006vx}.}. 
Many aspects of DBI scalar field theories have recently been discussed
in a number of papers \cite{Alishahiha:2004eh}-\cite{Bean:2007hc}, 
often assuming an $AdS_5$ throat ($f\sim1/\phi^4$) and various forms of
the potential $V$. 

The action \req{dbi} leads to field equations \req{conserv}, \req{friedman} 
for a perfect fluid with 
\bea
p     &=& \frac{\gamma-1}{f\gamma} - V(\phi) \label{pdbi}\\
\rho  &=& \frac{\gamma-1}{f} + V(\phi) \label{rhodbi}\ ,
\eea
where 
\be
\gamma = \frac{1}{\sqrt{1-f(\phi)\dot{\phi}^2}} \ .
\ee
To rewrite
the field equations in 
Hamilton-Jacobi form one proceeds as in the canonical case. From
\req{conserv}--\req{friedman} it follows that in this case that 
\bel{phidot}
\dot{\phi} = -\frac{2\mps}{\gamma} H'(\phi) \ .
\ee
This equation can easily be solved for $\dot{\phi}$ which allows one to 
express $\gamma$ as a function of $\phi$: 
\bel{gamma}
\gamma=\sqrt{1+ 4 M_P^4 f H'^2} \ .
\ee
Using this in \req{friedman} gives
\bel{hjdbi}
3\mps H^2 - V =  \frac{\gamma-1}{f} \ .
\ee
This is the Hamilton-Jacobi equation for DBI
inflation \cite{Silverstein:2003hf}. 

Using \req{pdbi} and \req{rhodbi} one finds
\be
p=\frac{\gamma^2-1}{f\gamma} - \rho \ .
\ee
Equating this to $w\rho$ and using  \req{gamma} and \req{hjdbi} gives
\bel{bardbi}
4\mps H'^2 = 3 (w+1) H^2 \sqrt{1 + 4 M_P^4 f H'^2} \ ,
\ee
which is the DBI analog of eq. \req{barcan}. This is a differential
equation for $H(\phi)$ which can in principle be solved for any prescribed 
$w(\phi)$. An analytic solution looks unlikely, but 
in the ``ultra-relativistic'' regime (considered
by \cite{Silverstein:2003hf}) this equation can easily be solved. 

The ``ultra-relativistic'' regime is that of large $\gamma$, where
one can use the approximation 
\be
\gamma \simeq 2\mps \sqrt{f} H' \ .
\ee
In this regime the Hamilton-Jacobi equation can be approximated by
\bel{hjuv}
3\mps H^2 - V = 2\mps \frac{H'}{\sqrt{f}}\ ,
\ee
and the scalar equation of motion simplifies to
\be
\dot{\phi} = - \frac{1}{\sqrt{f}} \ .
\ee
Returning to eq. \req{bardbi} the above approximations imply that 
one can neglect the one under the square root sign which leaves the simple
equation 
\bel{wdbi}
H' = \frac{3}{2} (w+1) \sqrt{f} H^2 \ .
\ee
The general solution of this equation can be written as
\bel{hdbisol}
H(\phi) = - \Big( C + \frac{3}{2} \int d\phi  (w(\phi)+1)
\sqrt{f(\phi)}\Big)^{-1}  \ ,
\ee
where $C$ is an integration constant. 

To find the corresponding potential one can use \req{hjuv}, which in
conjunction with \req{wdbi} leads to  
\bel{dbipot}
V(\phi) = - 3\mps w(\phi) H(\phi)^2 \ ,
\ee
where $H(\phi)$ is given by \req{hdbisol}. 

For constant $w$ the solution \req{hdbisol} is consistent with
\req{scalfact}, since the scalar field equation of motion in the
``ultra-relativistic'' regime is solved by 
\be
t = - \int d\phi \sqrt{f(\phi)} \ ,
\ee
so \req{hdbisol} yields 
\be
H(t) = \frac{2}{3(w+1)} \frac{1}{t-t_0} 
\ee
as implied by \req{scalfact}. In particular the number of n-folds can be
given as  
\be
N = \int dt H(t) = \frac{2}{3(w+1)} \ln (t_f/t_i) \ ,
\ee
where $t_i$ and $t_f$ refer to the cosmic time of the beginning and end of
inflation. All the dependence on the details of the model, i.e. on the
throat geometry and the potential, are hidden in $t_i$ and $t_f$ and the
constant value of $w$. 

Thus, for a given throat geometry there is always a choice of potential
which results in constant $w$ and therefore, in power law expansion. 
For the special case 
of $f(\phi)=\lambda/\phi^4$ 
considered in \cite{Silverstein:2003hf} one gets  
\bel{sth}
H(\phi) = \frac{2}{3\sqrt{\lambda}} \frac{1}{w+1}\frac{\phi}{1+C\phi} \ . 
\ee
This reduces to the well known DBI inflationary solution obtained in
\cite{Silverstein:2003hf} when $C=0$, or in the limit   
$\phi\rightarrow 0$ (which is independent of $C$ to leading order). In
\cite{Silverstein:2003hf} the authors assumed $H(\phi)=h_1\phi + \dots$ and
computed the 
corresponding potential, keeping in the end just the leading term
$V=V_2\phi$. One can easily verify that the condition on $V_2$ given in
\cite{Silverstein:2003hf} is just the condition $w<-1/3$ which ensures that
the expansion 
is accelerating. Indeed, for $C=0$ one gets (from \req{dbipot} and \req{sth})
\be
V(\phi) = - \frac{w}{(w+1)^2} \frac{4 \mps} {3 \lambda} \phi^2\ . 
\ee
The requirement that the inflaton mass be large for
``ultra-relativistic''  DBI inflation to be significant expresses the need
for $w$ to be close to $-1$.

\section{Conclusions}

DBI scalar field theories are an interesting and well motivated alternative 
to the usually considered models with canonical kinetic terms. While it is
possible that the higher derivatives of the inflaton field present in such
models are not relevant in practice and the usual slow roll scenario is
realized \cite{Shandera:2006ax,Spalinski:2007kt}, DBI models offer a new
possibility in the form of power 
law inflation with high inflaton mass \cite{Silverstein:2003hf}. In this
scenario the full DBI 
kinetic terms play an essential role. 
This possibility is interesting also because it is a way to obtain 
significant non-gaussianity in models of single field
inflation \cite{Alishahiha:2004eh}. It is expected that new data will tell 
whether this feature is attractive or not \cite{Lidsey:2006ia,Baumann:2006cd}.  
Note that the recent analysis carried out 
in reference \cite{Bean:2007hc} (see also \cite{Baumann:2006cd}) concluded 
that neither the KS  
throat nor any of the $Y^{p,q}$ throats can accommodate current
observational bounds on non-gaussianity when geometric constraints on
throat dimensions are applied. The results reported here may therefore be
useful in the study of alternative throat geometries if promising examples 
are found\footnote{ 
  References \cite{Baumann:2006cd},\cite{Bean:2007hc} suggest that
  orbifolds of the KS throat could be interesting in this context.}.
This note showed that power law inflationary 
solutions are possible for any throat geometry provided the potential 
is related to the throat function $f$ as discussed in the
previous section. The question of
whether power law inflation is a generic feature of 
DBI inflation will ultimately be resolved by string
theory computations of $f$ and $V$ in specific models.


\newpage
\begin{thebibliography}{00}

\bibitem{Dvali:1998pa}
  G.~R.~Dvali and S.~H.~H.~Tye,
  Phys.\ Lett.\  B {\bf 450}, 72 (1999)
  [arXiv:hep-ph/9812483].

\bibitem{Kachru:2003sx}
  S.~Kachru, R.~Kallosh, A.~Linde, J.~M.~Maldacena, L.~McAllister and
  S.~P.~Trivedi,
  JCAP {\bf 0310}, 013 (2003)
  [arXiv:hep-th/0308055].


\bibitem{Quevedo:2002xw}
  F.~Quevedo,
  Class.\ Quant.\ Grav.\  {\bf 19}, 5721 (2002)
  [arXiv:hep-th/0210292].


\bibitem{Cline:2006hu}
  J.~M.~Cline,
  ``String cosmology,''
  arXiv:hep-th/0612129.


\bibitem{Kallosh:2007ig}
  R.~Kallosh,
  ``On Inflation in String Theory,''
  arXiv:hep-th/0702059.

\bibitem{HenryTye:2006uv}
  S.~H.~Henry Tye,
  ``Brane inflation: String theory viewed from the cosmos,''
  arXiv:hep-th/0610221.

\bibitem{Leigh:1989jq}
  R.~G.~Leigh,
  Mod.\ Phys.\ Lett.\  A {\bf 4}, 2767 (1989).


\bibitem{Silverstein:2003hf}
  E.~Silverstein and D.~Tong,
  Phys.\ Rev.\  D {\bf 70}, 103505 (2004)
  [arXiv:hep-th/0310221].



\bibitem{Lucchin:1984yf}
  F.~Lucchin and S.~Matarrese,
  Phys.\ Rev.\  D {\bf 32}, 1316 (1985).

\bibitem{Becker:2005sg}
  K.~Becker, M.~Becker and A.~Krause,
  Nucl.\ Phys.\  B {\bf 715}, 349 (2005)
  [arXiv:hep-th/0501130].


\bibitem{Dimopoulos:2005ac}
  S.~Dimopoulos, S.~Kachru, J.~McGreevy and J.~G.~Wacker,
  ``N-flation,''
  arXiv:hep-th/0507205.

\bibitem{Barnaby:2004gg}
  N.~Barnaby, C.~P.~Burgess and J.~M.~Cline,
  JCAP {\bf 0504}, 007 (2005)
  [arXiv:hep-th/0412040].


\bibitem{Kofman:2005yz}
  L.~Kofman and P.~Yi,
  Phys.\ Rev.\  D {\bf 72}, 106001 (2005)
  [arXiv:hep-th/0507257].


\bibitem{Frey:2005jk}
  A.~R.~Frey, A.~Mazumdar and R.~Myers,
  Phys.\ Rev.\  D {\bf 73}, 026003 (2006)
  [arXiv:hep-th/0508139].


\bibitem{Chialva:2005zy}
  D.~Chialva, G.~Shiu and B.~Underwood,
  JHEP {\bf 0601}, 014 (2006)
  [arXiv:hep-th/0508229].


\bibitem{Allahverdi:2006iq}
  R.~Allahverdi, K.~Enqvist, J.~Garcia-Bellido and A.~Mazumdar,
  Phys.\ Rev.\ Lett.\  {\bf 97}, 191304 (2006)
  [arXiv:hep-ph/0605035].



\bibitem{Allahverdi:2007wh}
  R.~Allahverdi, A.~R.~Frey and A.~Mazumdar,
  ``Graceful exit from a stringy landscape via MSSM inflation,''
  arXiv:hep-th/0701233.



\bibitem{Giddings:2001yu}
  S.~B.~Giddings, S.~Kachru and J.~Polchinski,
  Phys.\ Rev.\  D {\bf 66}, 106006 (2002)
  [arXiv:hep-th/0105097].



\bibitem{Markov:1988yx} L.~P.~Grishchuk and Yu.~V.~Sidorov in 
  M.~A.~Markov, V.~A.~Berezin and V.~P.~Frolov,
  ``Quantum Gravity. Proceedings, 4TH Seminar, Moscow, USSR, May 25-29,
  1987'', {\it  World Scientiic (1988)}.


\bibitem{Muslimov:1990be}
  A.~G.~Muslimov,
  Class.\ Quant.\ Grav.\  {\bf 7}, 231 (1990).

\bibitem{Salopek:1990jq}
  D.~S.~Salopek and J.~R.~Bond,
  Phys.\ Rev.\  D {\bf 42}, 3936 (1990).

\bibitem{Kinney:1997ne}
  W.~H.~Kinney,
  Phys.\ Rev.\  D {\bf 56}, 2002 (1997)
  [arXiv:hep-ph/9702427].



\bibitem{Mukhanov:2005bu}
  V.~F.~Mukhanov and A.~Vikman,
  JCAP {\bf 0602}, 004 (2006)
  [arXiv:astro-ph/0512066].


\bibitem{Vikman:2006hk}
  A.~Vikman,
  ``Inflation with large gravitational waves,''
  arXiv:astro-ph/0606033.


\bibitem{Babichev:2006vx}
  E.~Babichev, V.~F.~Mukhanov and A.~Vikman,
  JHEP {\bf 0609}, 061 (2006)
  [arXiv:hep-th/0604075].



\bibitem{Alishahiha:2004eh}
  M.~Alishahiha, E.~Silverstein and D.~Tong,
  Phys.\ Rev.\  D {\bf 70}, 123505 (2004)
  [arXiv:hep-th/0404084].


\bibitem{Chen:2004hu}
  X.~Chen,
  Phys.\ Rev.\  D {\bf 71}, 026008 (2005)
  [arXiv:hep-th/0406198].


\bibitem{Chen:2005ad}
  X.~Chen,
  JHEP {\bf 0508}, 045 (2005)
  [arXiv:hep-th/0501184].


\bibitem{Kecskemeti:2006cg}
  S.~Kecskemeti, J.~Maiden, G.~Shiu and B.~Underwood,
  JHEP {\bf 0609}, 076 (2006)
  [arXiv:hep-th/0605189].




\bibitem{Shandera:2006ax}
  S.~E.~Shandera and S.~H.~Tye,
  JCAP {\bf 0605}, 007 (2006)
  [arXiv:hep-th/0601099].

\bibitem{Spalinski:2007kt}
  M.~Spali\'nski,
  JCAP {\bf 0704}, 018 (2007)
  [arXiv:hep-th/0702118].


\bibitem{Lidsey:2006ia}
  J.~E.~Lidsey and D.~Seery,
  Phys.\ Rev.\  D {\bf 75}, 043505 (2007)
  [arXiv:astro-ph/0610398].

\bibitem{Baumann:2006cd}
  D.~Baumann and L.~McAllister,
  ``A microscopic limit on gravitational waves from D-brane inflation,''
  arXiv:hep-th/0610285.

\bibitem{Bean:2007hc}
  R.~Bean, S.~E.~Shandera, S.~H.~Henry Tye and J.~Xu,
  ``Comparing brane inflation to WMAP,''
  arXiv:hep-th/0702107.



\end{thebibliography}
\end{document}